\documentstyle[12pt,epsfig]{article}

\advance\textheight by \topskip
\begin{document}
\tolerance=100000
\thispagestyle{empty}
\setcounter{page}{0}

\newcommand{\be}{\begin{equation}}
\newcommand{\ee}{\end{equation}}
\newcommand{\br}{\begin{eqnarray}}
\newcommand{\er}{\end{eqnarray}}
\newcommand{\ba}{\begin{array}}
\newcommand{\ea}{\end{array}}
\newcommand{\bi}{\begin{itemize}}
\newcommand{\ei}{\end{itemize}}
\newcommand{\bn}{\begin{enumerate}}
\newcommand{\en}{\end{enumerate}}
\newcommand{\bc}{\begin{center}}
\newcommand{\ec}{\end{center}}
\newcommand{\ul}{\underline}
\newcommand{\ol}{\overline}
\def\epem{\ifmmode{e^+ e^-} \else{$e^+ e^-$} \fi}
\newcommand{\eeww}{$e^+e^-\rightarrow W^+ W^-$}
\newcommand{\qqQQ}{$q_1\bar q_2 Q_3\bar Q_4$}
\newcommand{\eeqqQQ}{$e^+e^-\rightarrow q_1\bar q_2 Q_3\bar Q_4$}
\newcommand{\eewwqqqq}{$e^+e^-\rightarrow W^+ W^-\ar q\bar q Q\bar Q$}
\newcommand{\eeqqgg}{$e^+e^-\rightarrow q\bar q gg$}
\newcommand{\eeqloop}{$e^+e^-\rightarrow q\bar q gg$ via quark loops}
\newcommand{\eeqqqq}{$e^+e^-\rightarrow q\bar q Q\bar Q$}
\newcommand{\eewwjjjj}{$e^+e^-\rightarrow W^+ W^-\rightarrow 4~{\rm{jet}}$}
\newcommand{\eeqqggjjjj}{$e^+e^-\rightarrow q\bar 
q gg\rightarrow 4~{\rm{jet}}$}
\newcommand{\eeqloopjjjj}{$e^+e^-\rightarrow q\bar 
q gg\rightarrow 4~{\rm{jet}}$ via quark loops}
\newcommand{\eeqqqqjjjj}{$e^+e^-\rightarrow q\bar q Q\bar Q\rightarrow
4~{\rm{jet}}$}
\newcommand{\eejjjj}{$e^+e^-\rightarrow 4~{\rm{jet}}$}
\newcommand{\jjjj}{$4~{\rm{jet}}$}
\newcommand{\qqbar}{$q\bar q$}
\newcommand{\ww}{$W^+W^-$}
\newcommand{\ar}{\rightarrow}
\newcommand{\sm}{${\cal {SM}}$}
\newcommand{\Dir}{\kern -6.4pt\Big{/}}
\newcommand{\Dirin}{\kern -10.4pt\Big{/}\kern 4.4pt}
\newcommand{\DDir}{\kern -7.6pt\Big{/}}
\newcommand{\DGir}{\kern -6.0pt\Big{/}}
\newcommand{\wwqqqq}{$W^+ W^-\ar q\bar q Q\bar Q$}
\newcommand{\qqgg}{$q\bar q gg$}
\newcommand{\qloop}{$q\bar q gg$ via quark loops}
\newcommand{\qqqq}{$q\bar q Q\bar Q$}

\def\l{\left\langle}
\def\r{\right\rangle}
\def\aem{\alpha_{\rm em}}
\def\as{\alpha_{\rm s}}
\def\MW{M_{W^\pm}}
\def\MZ{M_{Z}}
\def\ycut{y_{\rm cut}}
\def\Ord{\lower .7ex\hbox{$\;\stackrel{\textstyle <}{\sim}\;$}}
\def\OOrd{\lower .7ex\hbox{$\;\stackrel{\textstyle >}{\sim}\;$}}
\def\pl #1 #2 #3 {{\it Phys.~Lett.} {\bf#1} (#2) #3}
\def\np #1 #2 #3 {{\it Nucl.~Phys.} {\bf#1} (#2) #3}
\def\jp #1 #2 #3 {{\it J.~Phys.} {\bf#1} (#2) #3}
\def\zp #1 #2 #3 {{\it Z.~Phys.} {\bf#1} (#2) #3}
\def\pr #1 #2 #3 {{\it Phys.~Rev.} {\bf#1} (#2) #3}
\def\prep #1 #2 #3 {{\it Phys.~Rep.} {\bf#1} (#2) #3}
\def\prl #1 #2 #3 {{\it Phys.~Rev.~Lett.} {\bf#1} (#2) #3}
\def\mpl #1 #2 #3 {{\it Mod.~Phys.~Lett.} {\bf#1} (#2) #3}
\def\rmp #1 #2 #3 {{\it Rev. Mod. Phys.} {\bf#1} (#2) #3}
\def\sjnp #1 #2 #3 {{\it Sov. J. Nucl. Phys.} {\bf#1} (#2) #3}
\def\cpc #1 #2 #3 {{\it Comp. Phys. Commun.} {\bf#1} (#2) #3}
\def\xx #1 #2 #3 {{\bf#1}, (#2) #3}
\def\preprint{{\it preprint}}

\begin{flushright}
{\large RAL-TR-98-036}\\
{\large DTP/98/14}\\
{\rm June 1999\hspace*{.5 truecm}}\\
\end{flushright}

\vspace*{\fill}

\begin{center}
{\Large \bf Hadronic returns to the ${\boldmath Z}$ \\[3mm]
in electron-positron annihilation at high energy}\\[1.cm]
{\large V.A.~Khoze$^{a,b,}$\footnote{E-mails:
Valery.Khoze@vxcern.cern.ch;
Moretti,D.J.Miller@rl.ac.uk;
W.J.Stirling@durham.ac.uk.}, D.J. Miller$^{c,1}$,
S.~Moretti$^{c,1,}$
and W.J.~Stirling$^{a,d,1}$}\\[0.4 cm]
{\it a) Department of Physics, University of Durham,}\\
{\it South Road, Durham DH1 3LE, UK.}\\
{\it b) INFN--Laboratori Nazionali di Frascati,}\\
{\it P.O. Box 13, I-00044 Frascati (Rome), Italy.}\\
{\it c) Rutherford Appleton Laboratory,}\\
{\it Chilton, Didcot, Oxon OX11 0QX, UK.}\\
{\it d) Department of Mathematical Sciences, University of Durham,}\\
{\it South Road, Durham DH1 3LE, UK.}\\
\end{center}

\vspace*{\fill}

\begin{abstract}
{\small
\noindent
The production of four hadronic jets in $e^+e^-$ collisions above
the $Z$ pole is dominated by the QCD $e^+e^- \to q\bar q q \bar q,q\bar q gg$
processes and, for sufficiently  high energy,  the electroweak
$e^+e^- \to W^+W^- \to q\bar q q \bar q$ process. However there is another
mechanism for producing four jets, $e^+e^- \to Z gg \to q\bar q gg$,
which can be regarded as a ``hadronic return'' to the $Z$ pole. We
investigate
this new process in detail.}
\end{abstract}

\vspace*{\fill}
\newpage

\section{Introduction and motivation}
\label{sec_intro}

For $e^+e^-$ centre-of-mass (CM) 
collision energies above the $Z$ pole, $\sqrt{s} > M_Z$, 
an important contribution to the cross section for $e^+e^- \to 
Z \to $~jets comes from the  so-called 
``photonic returns
to the $Z$'', i.e events in which  one
or more photons are emitted by the colliding electron-positron beams
prior to the annihilation of the latter into the virtual $Z$ \cite{photonthe}.
Such emission can carry away a large fraction of the collider energy, 
so that the amount left over for the effective $e^+e^-$ interaction,
$\sqrt s_{\mathrm{eff}}$, can resonate at $\sqrt s_{\mathrm{eff}} \approx
M_Z$,  well below the nominal value of
$\sqrt s$. Furthermore, such  initial state radiation (ISR)
is dominantly  collinear to the beam direction, and so many of the photons
are in practice not seen by the detectors.
At LEP2, with $ \sqrt s \Ord~200$ GeV,  hadronic events 
in which $\sqrt s_{\mathrm{eff}}\approx M_Z$ have indeed been observed
and in fact, 
because of the resonant shape of the $e^+e^-$ cross 
section in the vicinity of the $Z$ pole, they
constitute a large fraction (about one quarter)
of the total $e^+e^-$ production rate. If one wants to
study ``genuine'' physics at the LEP2 collision energy scale
one has to remove such events from the analysis. This is also a concern
for a future electron-positron linear collider (NLC), where the
effect of ISR is a subject of ongoing study \cite{ISR}.

\begin{figure}[p]
\begin{center}
\vskip-5.0cm
~\epsfig{file=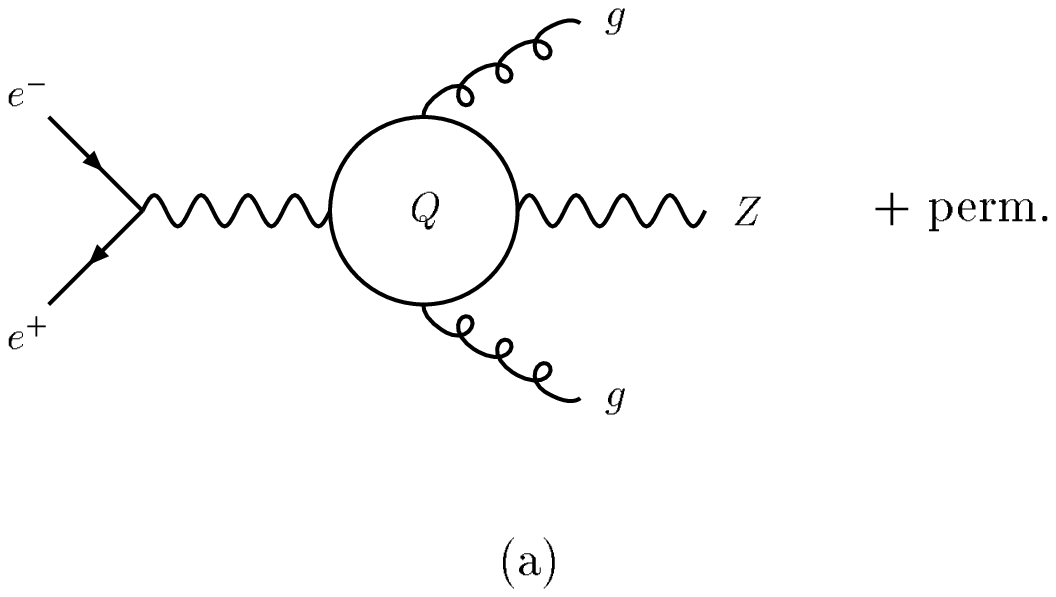,height=18cm}
\vskip-14.0cm
~\epsfig{file=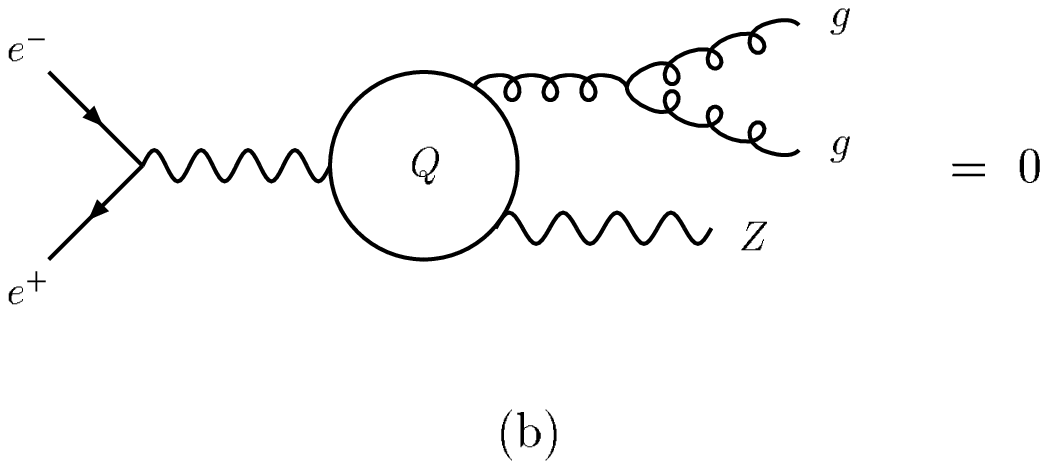,height=18cm}
\vskip-12.75cm
~\epsfig{file=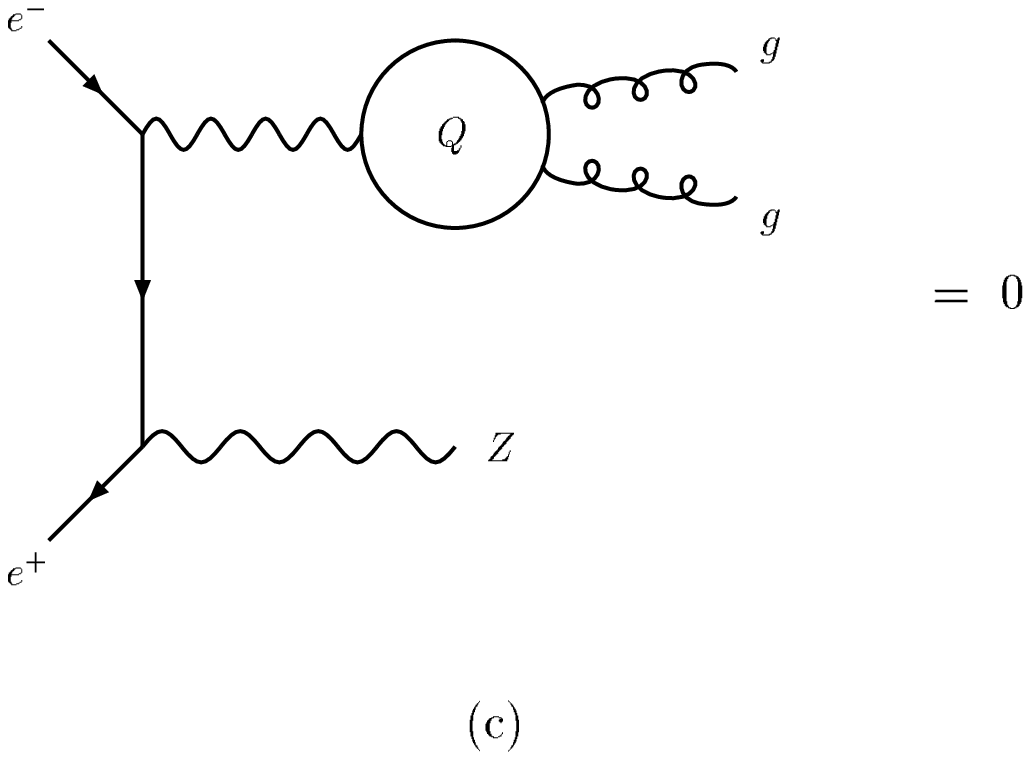,height=18cm}
\vskip-10cm
\caption{Lowest order diagrams responsible for the $e^+e^-\ar Zgg$
process. 
An internal wavy line represents a $\gamma$ or a $Z$. 
Permutations are not shown. Graphs (a) are the hadronic returns.
Graphs (b) and (c) are prohibited, as discussed in the text.}
\label{fig_Zgg}
\end{center}
\end{figure}

There is  another interesting way to return the collider CM energy to the
vicinity of $M_Z$  in $e^+e^-$ annihilation. It proceeds 
via strong (QCD) rather than electromagnetic (EM) interactions.
The mechanism is illustrated schematically
in Fig.~\ref{fig_Zgg}a. The virtual $Z$ (or photon) fluctuates
into a quark-antiquark loop from which two energetic gluons are emitted.
Note that the box diagrams are the
only ones that contribute at this leading 
 ${\cal O}(e^3g_{\mathrm{s}}^2)$ order. Triangle diagrams in which
the two external gluons are produced via a triple-gluon vertex
(see Fig.~\ref{fig_Zgg}b) are identically zero 
by colour conservation,  and triangle diagrams
involving a $\gamma,Z$ splitting into a pair of on-shell gluons (see
Fig.~\ref{fig_Zgg}c) are forbidden by the Landau-Yang's theorem.  
Thus the emission of two gluons from an internal
quark loop  allows for a reduction of the incoming $\sqrt s$
energy of the $\gamma,Z$ current into a smaller outgoing energy 
$\sqrt s_{\mathrm{eff}}$ of order $M_Z$, such that an on-shell neutral 
electroweak (EW) gauge
boson can indeed materialise. We may therefore call such events
 ``hadronic returns to the $Z$'', or HR for short.
 
At first sight, the HR cross section would appear
to be heavily suppressed compared to the standard EM
return. For example, counting powers of coupling constants in the 
matrix elements squared shows that the former
 is of order
${\cal O}(\alpha_{\mathrm{em}}^3 \alpha_{\mathrm{s}}^2)$ whereas
the latter is of order ${\cal O}(\alpha_{\mathrm{em}}^2)$ only.
HR events  should therefore be a factor of order $10^4$ less 
frequent than   photonic return events. Nevertheless it is worth investigating
whether the HR events are observable at all at LEP2 and/or NLC.
If one continues the simple exercise
of coupling counting  and multiplies the $e^+e^-$ production cross
section at the $Z$ peak, which is of order ${\cal O}(\alpha_{\mathrm{em}})$,
times $\alpha_{\mathrm{em}}^2 \alpha_{\mathrm{s}}^2$,
one  obtains an estimate for  the HR
cross section of the order of a few events
per hundred inverse picobarn of luminosity  at either machine. 
Since the typical luminosity sample expected at the end of the
LEP2 running period is at least of this order of magnitude,  HR events 
may well be already observable.
For the NLC,  the figure currently foreseen for the yearly luminosity
is of the order of 100~fb$^{-1}$,  which could correspond to
hundreds of HR events.

Furthermore, the kinematics of HR events is rather peculiar. As no infrared
singularities exist in the loop tensor
associated with the double gluon emission
(see Ref.~\cite{Laursen} for a discussion), one expects 
a clear ``$Z+2$jet'' signal, with the 
two gluon jets being  energetic
and quite randomly distributed in the relative angle. The jet--jet invariant
mass distribution should be broad, without the low-mass
peaking associated with $g^* \to gg$ for example. 
In this respect, the HR events could be a sizeable background
for processes like $e^+e^- \to W^+W^-\to q\bar q' Q\bar Q'$ and
$e^+e^- \to ZH \to q\bar q b \bar b$,  the latter 
when the $Z$ and $H$ masses are degenerate. In both cases,
 all jets are naturally energetic and well separated. Furthermore,
in the second example, 
the resonant $bb$ pair  in $Zgg$ events is furnished by the decaying gauge  
boson, mimicking  $H\to b\bar b$.

Of course HR is not the only source of $Z+2$jet events in the Standard
Model. The main background (assuming that the $Z$ is clearly identified,
for example via its leptonic decays) comes from the 
${\cal O}(\alpha_{\mathrm{em}}^3 )$ $e^+e^- \to Z q \bar q$ process.
But here the jet--jet mass distribution is strongly peaked at
$M_{jj} \sim 0$ and also, if kinematics allow, at $M_Z$.
 A quantitative comparison with HR events will be presented below.
If the $Z$ boson in HR events decays to two jets, there is 
of course  a large background from standard QCD and EW
$2\ar 4$ processes \cite{twotofour}. However the topology
of such events is in general very different from the HR one, as we shall see.

The discussion so far has been at the level
of  coupling constants and general
kinematics. That, however, is only part of the story.
To quantify the above effects one must calculate the appropriate
one-loop Feynman diagrams.
In fact, it is very difficult to make any {\sl a priori} statements
regarding the loop dynamics.
Although the expression for the tensor box
entering in Fig.~\ref{fig_Zgg}a, with two massive and two massless
external legs, has been known for a long time \cite{tensor}\footnote{It has
also been  
reproduced in Refs.~\cite{nigel,jochum}.} those results cannot be
used for the HR cross section calculation here. This is due to the presence of a
{\sl Gram determinant} that causes problems with numerical stability when the
matrix elements are interfaced with the $1 \to 3$ phase space. 
This will be discussed further below.

The paper is organised as follows.
In Section~\ref{sec_calculation} we outline how the calculation is
performed and make use of the results in order to assess whether hadronic
returns induced by the diagrams in Fig.~\ref{fig_Zgg}a can be of any relevance
at all in phenomenological analyses at present and/or future
electron-positron colliders. The results presented in Section~\ref{sec_results}
will help answer this question. In Section~\ref{sec_conclusions} we will
summarise the main findings of our studies. 
  
\section{The calculation}
\label{sec_calculation}

As mentioned in the introduction, the fourth-rank tensor that enters the
amplitude squared for $e^+e^-\ar Zgg$
associated with the (six) diagrams in Fig.~\ref{fig_Zgg}a requires special
attention in order to resolve problems with its numerical stability.

As with all calculations at one-loop order, explicit analytic formulae contain
determinants in the denominator of the expressions. These determinants can be
thought of as arising from the inversion of a set of simultaneous equations,
and can be, in general, complicated polynomials of the invariants of the
problem. Furthermore, these determinants are raised to a power of up to the  
rank of the tensor (i.e. in our case, four). Clearly, any analytic
expression will be numerically unstable close to the point in phase space
where this determinant vanishes. This is the Gram determinant stability
problem~\cite{gram}. 

In most previous calculations involving triangle and box diagrams this was not
significant because the Gram determinant vanished only at the edge of phase
space, where the matrix elements diverged anyway due to soft and/or
collinear singularities. Indeed this is the case in
Refs.~\cite{tensor,nigel,jochum} where the box diagrams were only required
for $2 \to 2$ processes, ensuring that the determinant singularity was safely
stowed away at the edge of phase space. However, when one considers the same
box integral for $1 \to 3$ processes, as we must do here, the Gram determinant
singularity now falls in the centre of our  phase space and cannot be
ignored. 

It should be emphasised that this is a purely technical, rather then a
conceptual, problem. There is, in actuality, no physical divergence when the
Gram determinant vanishes. It is, in fact, merely an artifact of the way in
which the calculation was performed, where the expression is written in a form
which makes a {\sl fake} divergence apparent. That is, it should be possible
to overcome the Gram determinant problem by either performing the numerical
integration at a sufficiently high accuracy or by analytically cancelling
by hand divergent terms in the expression to render the matrix elements
manifestly finite. However, in the tensor integral of
Refs.~\cite{nigel,jochum}, the Gram determinant in the denominator is raised
to the fourth power, giving a divergence which is too powerful to overcome
even at quadruple precision. Using this tensor for the $1 \to 3$ process
one would obtain wildly inaccurate results which could not be trusted.
Instead
one must adopt the second approach and write the matrix elements in a
manifestly finite form.

This is not as easy as it sounds. The one-loop matrix elements are extremely
complicated and lengthy, containing logarithms, dilogarithms and
polynomials of the invariants. It is virtually impossible to find and cancel
the divergent terms in such an expression. Instead, we take the approach
of Ref.~\cite{gram}, and recalculate the tensor box integral in such a way that
these fake Gram determinant singularities are controlled {\sl from the beginning}.
In this way we have recalculated the tensor box integral such that it is
manifestly finite over all phase space and can be used for $1 \to 3$ processes
without any problems of stability. This method also leads to more compact
expressions (although still too lengthy to be reproduced here). 

The newly calculated tensor is then interfaced with the incoming $e^+e^-\ar
\gamma^*,Z^*$ off-shell current and the outgoing polarisation vectors for the
gluons and the gauge boson. In some cases we have kept the latter off-shell,
also  allowing for contemporaneous $\gamma^* gg$ and $Z^* gg$ production 
followed by the decays $\gamma^*,Z^*\ar f\bar f$.

The $2\ar3$ and $2\ar4$ matrix elements (MEs) obtained this way have been
integrated numerically over the appropriate three- and four-body phase spaces.
Because of delicate cancellations taking place among the diagrams,
it is however important to verify the
stability of the results against different mappings of the latter.
To do so we have implemented the kinematics of the final states
both analytically, using different integration variables
(see, for example, Ref.~\cite{Kajantie} for some possible choices),
and numerically
using the multi-particle phase space generator {\tt RAMBO} \cite{RAMBO}.
Furthermore, different routines have been used for the numerical integration,
namely the Monte Carlo package {\tt VEGAS} \cite{VEGAS}
and the {\tt NAGLIB} routines
{\tt D01EAF} and {\tt D01GCF}. We have found agreement within the
numerical errors for all the resulting outputs.
An optimised version of the program designed  for high statistics Monte Carlo
simulations is available upon request from the authors.

The numerical values  for the electroweak and strong parameters used
in the numerical calculations presented below are
as follows: $\sin^2\theta_W=0.2320$, $M_Z=91.19$~GeV, $\Gamma_Z=2.5$~GeV,
$\MW\equiv\MZ\cos\theta_W\approx80$~GeV, $\Gamma_{W^\pm}=2.2$~GeV,
$\aem= 1/128$ and $\as$ is computed at two-loop, with
five active flavours and at a renormalisation scale equal to the CM energy.
All fermion (i.e. lepton and five quark) masses 
are set equal to zero. 
As  representative of LEP2 and NLC we have considered 
CM energies
in the range 130~GeV $\Ord\sqrt s\equiv E_{\mathrm{cm}}\Ord$ 500~GeV.
Given the value of the top mass, i.e. $m_t \approx 175$ GeV, and the presence
of four top propagators in the box diagrams of Fig.~\ref{fig_Zgg}a, the
contribution of the top loop to the overall HR
cross section is negligible, both at LEP2 and NLC, and
so we do not include it here for simplicity\footnote{A similar remark 
applies to the top loop contribution in 
the  process $gg\ar ZZ$ at LHC \cite{nigel}. In addition, a
non-zero mass in the loop leads to a large proliferation of terms
in the tensor reduction, see Ref.~\cite{jochum}.}.

\section{Results}
\label{sec_results}

\begin{figure}[p]
~\epsfig{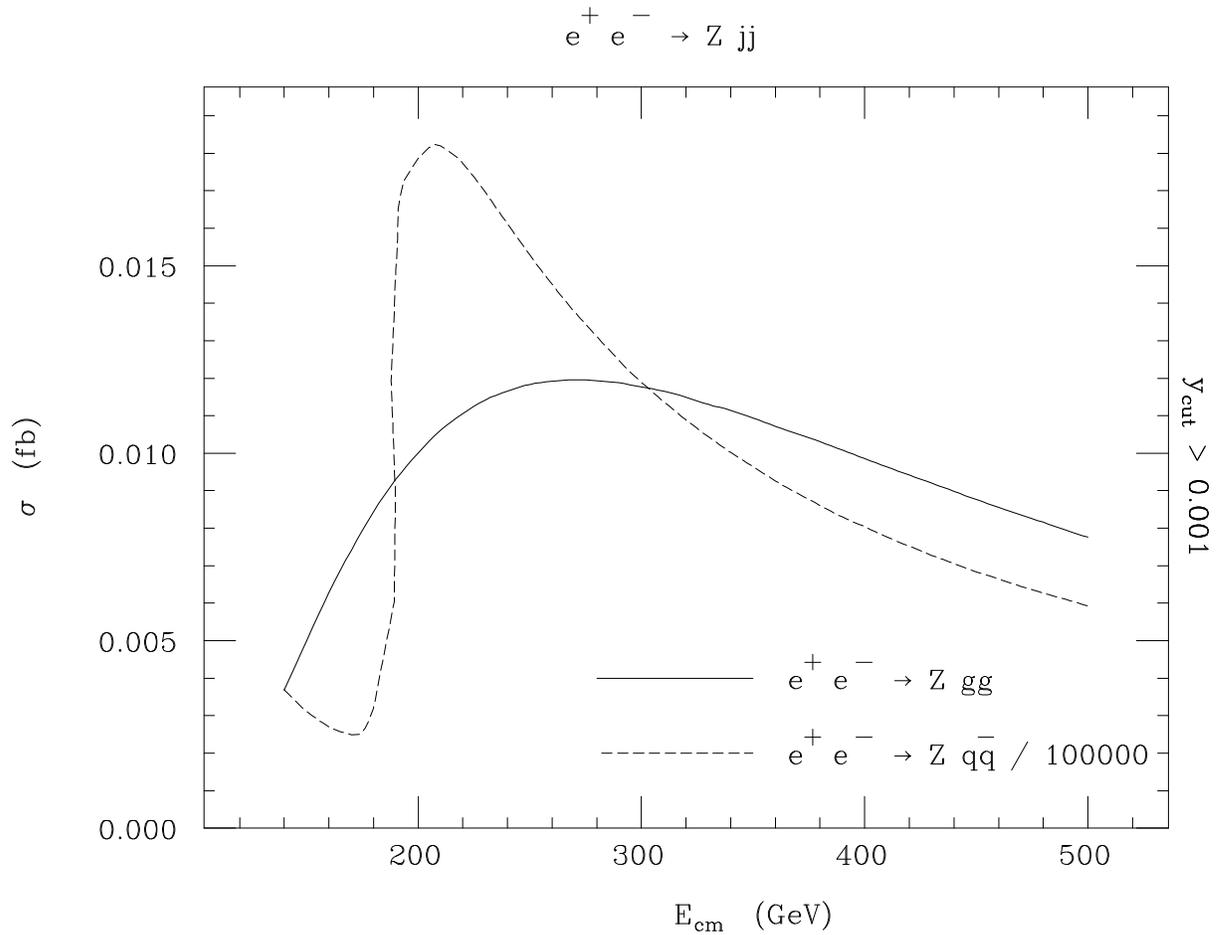}
\vspace*{-1cm}
\caption{Cross section for $e^+e^-\ar Zgg$ events as a function of
the CM energy. The jet clustering algorithm 
used to separate the gluon jets is the Durham algorithm, 
with cut-off $y_{\mathrm{cut}}>0.001$. Also shown is the EW $e^+e^-
\ar Z q \bar q$ cross section, scaled by a factor of $10^{-5}$.}
\label{fig_scan}
\end{figure}

We start our investigation of the hadronic returns to the $Z$ by studying
the production cross section of the process $e^+e^-\ar Zgg$ as a function of 
$E_{\mathrm{cm}}$, between typical LEP2 and NLC energies. 
In order to observe the two gluons as separate jets, one ought to impose
some isolation criteria on them. To this end, we simply adopt a jet clustering
algorithm \cite{schemes}: for example, the Durham algorithm \cite{DURHAM}, 
with resolution $y_{\mathrm{cut}}=0.001$. (Note however that none of the main
features of our analysis depends upon either the choice of the jet
algorithm or of its resolution parameter.)
The curve in Fig.~\ref{fig_scan} reports the production rates with 
such a di-jet selection enforced.

We see from Fig.~\ref{fig_scan} that $\sigma(e^+e^-\ar Zgg)$ is very small.
The maximum value occurs at $E_{\mathrm{cm}}\approx 280$ GeV,
and this is of the order of 0.012 fb only.
For typical LEP2 energies, it is always smaller than 0.010~fb, well 
below detection level. In fact, at least 100 inverse femtobarns of LEP2 luminosity
would need to be collected  to observe just one such event, a  figure which 
is well beyond the  current machine potential.
At the NLC, running at around the top-antitop threshold, i.e. 
with $E_{\mathrm{cm}}\approx2m_t$, where $m_t=175$ GeV, the production rate
is slightly above 0.010 fb. For 100 fb$^{-1}$ per
year, a handful of $e^+e^-\ar Zgg$ events would be produced
at the end of the collider lifetime at such energy (say, of about five years).
At an NLC running at 500~GeV the rates fall back to the values
typical of LEP2. However, the order of magnitude of yearly luminosity
of a linear collider running with $E_{\mathrm{cm}}\gg2m_t$ is 
expected to be not
much different from the corresponding value at threshold. Thus
some HR events would eventually show up also at an NLC
running at high energy. 

Figure~\ref{fig_scan} also shows the cross section for the 
${\cal O}(\alpha_{\mathrm{em}}^3 )$ $e^+e^- \to Z q \bar q$ process.
This is {\sl much} larger than the HR cross section, but the kinematics
are very different. In fact at CM energies greater than about
$2 M_Z$ the $Z q \bar q$ cross section is completely dominated
by the on-shell $ZZ \to Z q \bar q$ contribution, corresponding
to final states with $M_{jj} \sim M_Z$. We shall return to this below when we
consider kinematic distributions at a typical NLC energy.

\begin{figure}[p]
~\epsfig{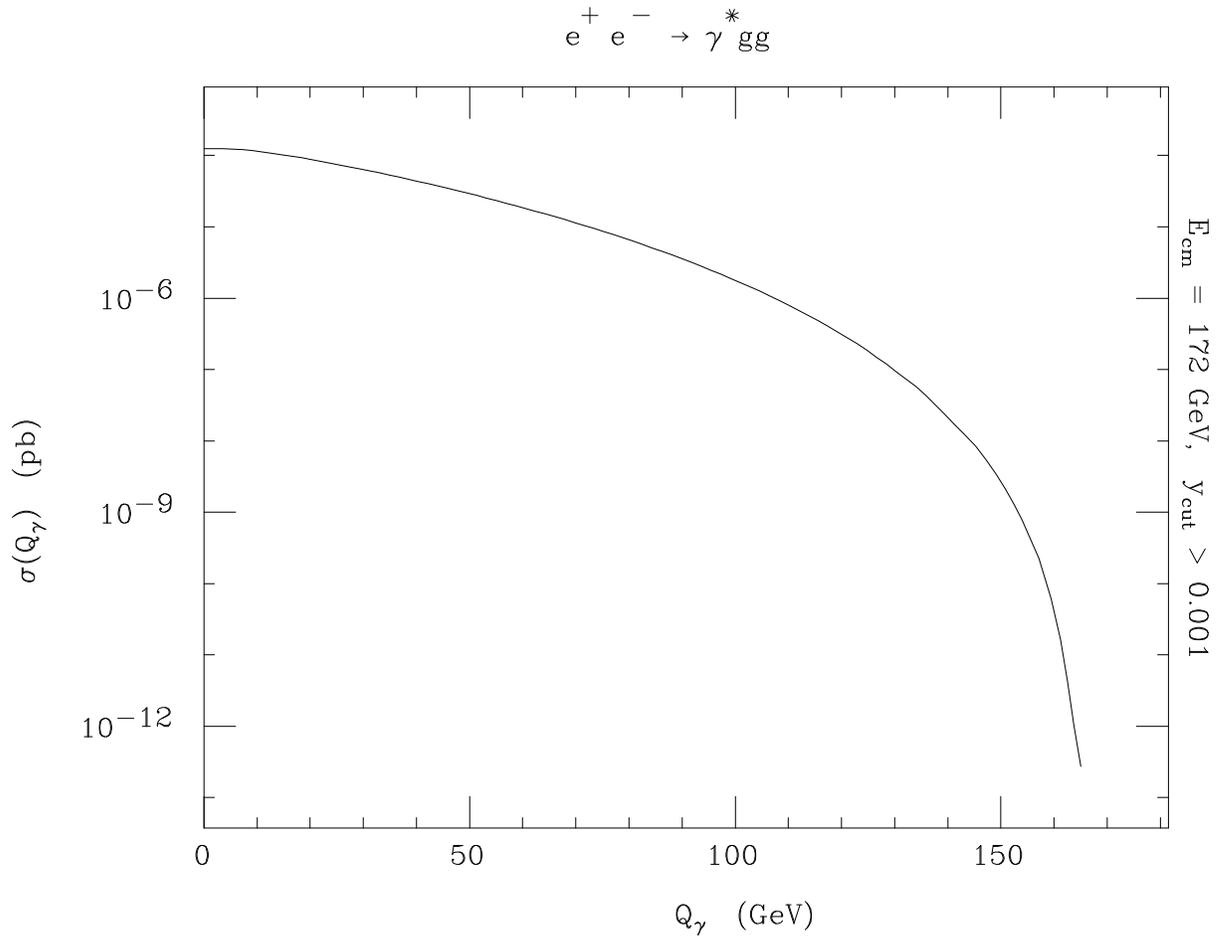}
\vspace*{-1cm}
\caption{Cross section for $e^+e^-\ar \gamma^* gg$ events as a function 
of the photon virtuality $Q_\gamma$ at $E_{\mathrm{cm}}=172$~GeV.
The jet clustering algorithm 
used to separate the gluon jets is the Durham one, 
with cut-off $y_{\mathrm{cut}}>0.001$.}
\label{fig_Mqq}
\end{figure}

The HR cross section values in Fig.~\ref{fig_scan} seem to contradict the speculations
made in the introduction, where we had argued in terms of couplings
and kinematics about the possible production rate 
of $e^+e^-\ar Zgg$ events. In fact our original (optimistic) arguments there are
spoiled by the loop behaviour --- the far-off-shell internal propagators
are an important additional  source of suppression.
To understand this further 
we show in Fig.~\ref{fig_Mqq} the cross section for $e^+e^-\ar \gamma^*gg$,
i.e. for the production of an off-shell photon of virtuality $Q_\gamma$,
as a function of the latter. For illustrative purposes, we
consider a CM energy of 172~GeV. Notice how the cross section decreases
by almost two orders of magnitude going from $Q \approx 0$ to $Q \approx M_Z$.
Our arguments about the size of the cross section based on counting
powers of the coupling did not take this $Q$ dependence into account, 
and in fact they turn out to be  relevant for the small $Q$ limit only. 
The decrease of available
phase space as $Q$ increases is another important factor at this energy.   

\begin{figure}[p]
~\epsfig{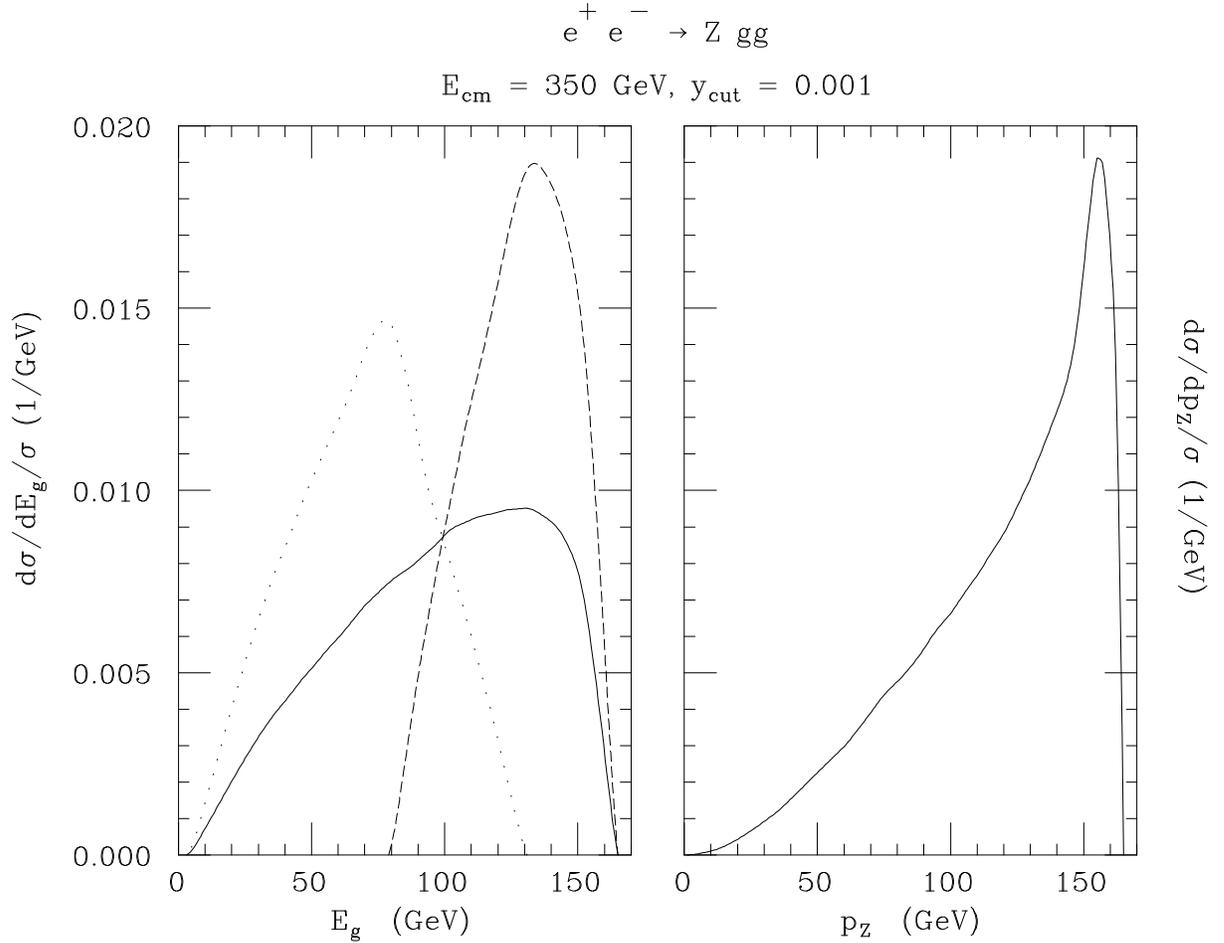}
\vspace*{-1.0cm}
\caption{Differential distributions  in the gluon energy (left:
solid for any of the gluons and dashed[dotted] for the most[least] energetic
one) and in the $Z$ momentum (right)
for $e^+e^-\ar Zgg$ events at $E_{\mathrm{cm}}=350$~GeV. 
The jet clustering algorithm 
used to separate the gluon jets is the Durham algorithm, 
with cut-off $y_{\mathrm{cut}}>0.001$.
Normalisation is to unity.}
\label{fig_Zgg1}
\end{figure}

\begin{figure}[p]
~\epsfig{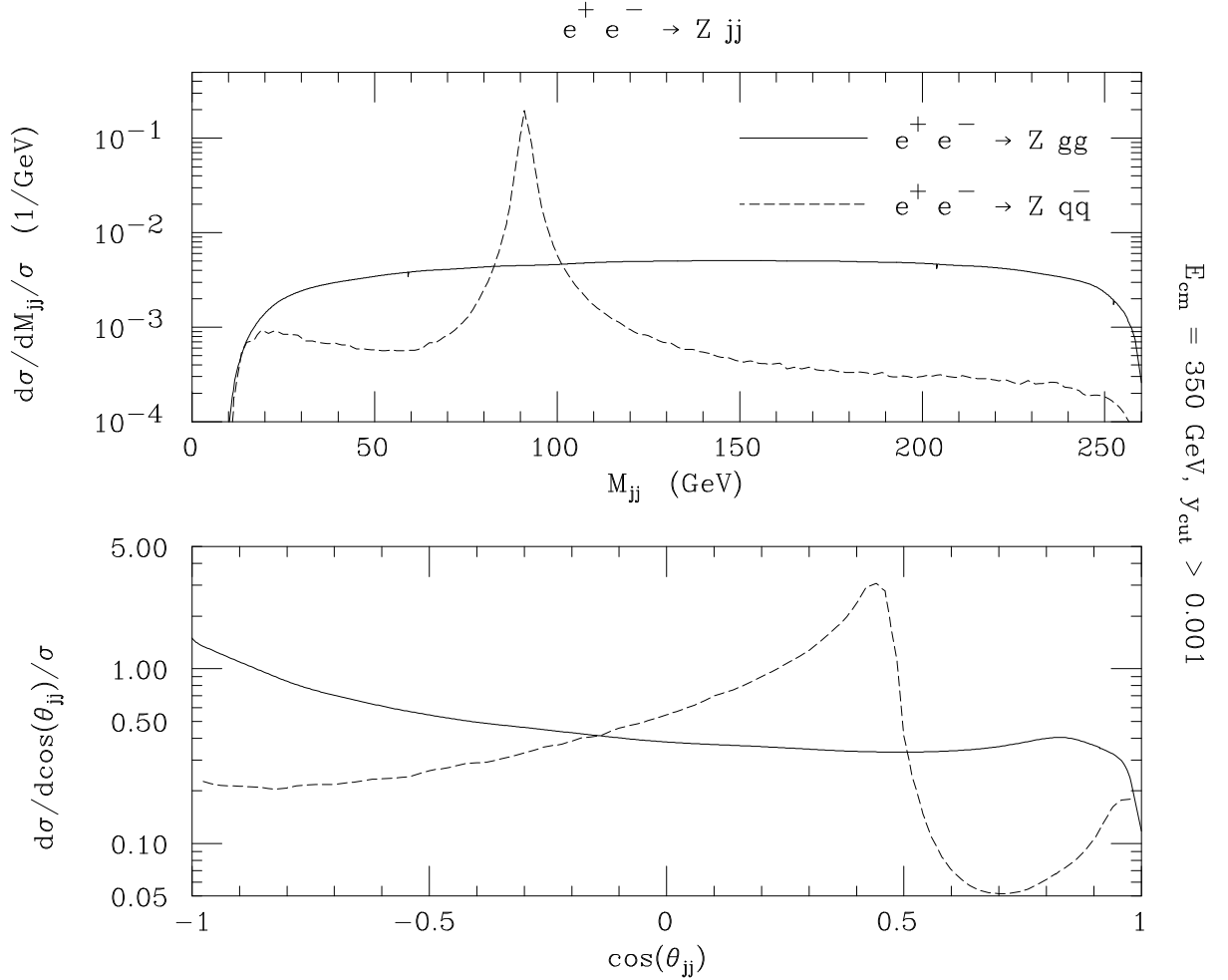}
\vspace*{-1cm}
\caption{Differential distributions in the gluon-gluon invariant mass
(upper plot) and relative cosine (lower plot) 
for $e^+e^-\ar Zgg$ events at $E_{\mathrm{cm}}=350$~GeV. 
The jet clustering algorithm 
used to separate the gluon jets is the Durham algorithm, 
with cut-off $y_{\mathrm{cut}}>0.001$.
Normalisation is to unity. Also shown are the corresponding
distributions for the EW $e^+e^-
\ar Z q \bar q$ process.}
\label{fig_Zgg2}
\end{figure}

Having understood the origin of the overall size of the production
cross section for $e^+e^-\ar Zgg$, we turn  to
some typical kinematic distributions of such events at the NLC.
The typical energies (momenta)
of the two gluons can be seen in the left-hand
plot of Fig.~\ref{fig_Zgg1}, whereas in the right-hand plot of the same
figure we show the distribution in the 
momentum of the $Z$-boson. In Fig.~\ref{fig_Zgg2}, we
show instead the invariant mass of the two gluons  (upper plot),
together with the angular separation between them (lower plot). 
The distributions shown in these two figures agree 
 remarkably well with the behaviour anticipated in
Sect.~\ref{sec_intro}. That is, 
the two gluons are produced with large energy and with no tendency 
to be  emitted in the same direction (associated with the absence 
of soft and collinear singularities in the ME). Indeed, in many events
the gluons are approximately 
back-to-back (see lower plot of Fig.~\ref{fig_Zgg2}). 
Fig.~\ref{fig_Zgg2} also shows the invariant mass and angular distributions
for the $Z q \bar q$ process. As anticipated, the former contains a sharp
Breit-Wigner peak at $M_{jj} \sim M_Z$ which is responsible for the bulk
of the cross section. The same kinematics are also responsible for the 
jets being produced at a relative angle  of about $60^\circ$. 

\begin{figure}[p]
~\epsfig{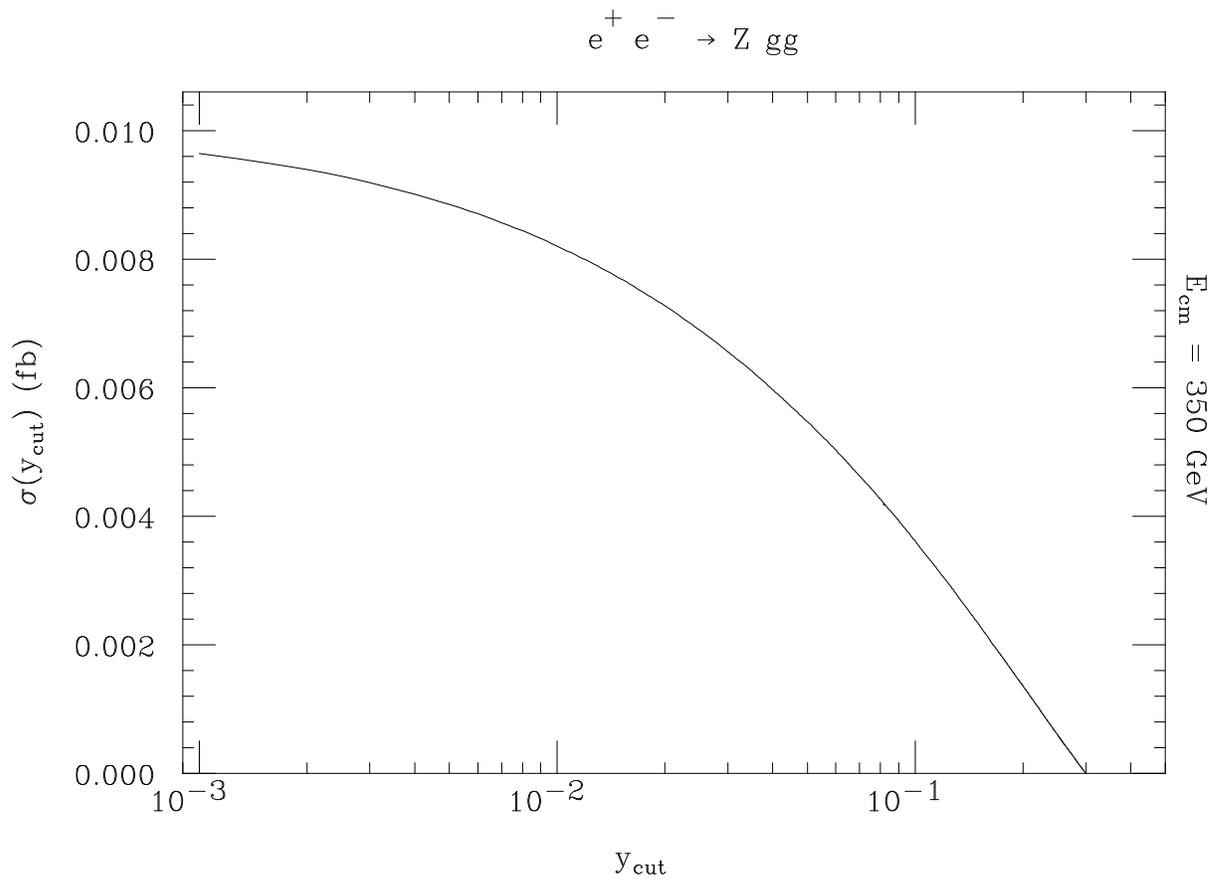}
\vspace*{-1.0cm}
\caption{HR cross section 
as a function of the cut-off $y_{\mathrm{cut}}$ in the Durham jet algorithm
for $e^+e^-\ar Zgg$ events at $E_{\mathrm{cm}}=350$~GeV.}
\label{fig_Zgg3}
\end{figure}

In Figs.~\ref{fig_Zgg1}--\ref{fig_Zgg2} the default value of 0.001
was used for the resolution parameter $y_{\mathrm{cut}}$ of the Durham jet algorithm.
However, given the definition of the separation
\begin{equation}\label{D} 
y_{ij} = {{2\min (E^2_i, E^2_j)(1-\cos\theta_{ij})}\over{s}}
\end{equation}
which has to be compared to $y_{\mathrm{cut}}$,
in terms of the energies $E_i$ and $E_j$ and relative angle
$\theta_{ij}$ for each pair of particles $ij$ (here, the $gg$ pair only),
the kinematics described above  imply that there is in fact  very 
little dependence of $\sigma(e^+e^-\ar Zgg)$ on the actual value of
$y_{\mathrm{cut}}$. This is illustrated in Fig.~\ref{fig_Zgg3}.
If one increases the cut-off by, say, a factor of ten from the default
value (0.001), approximately $15\%$ more events are rejected.  

So far we have only considered the case of {\sl on-shell} $Z$ boson
production. In other words, the rates given
so far correspond to what one would obtain by adding together 
all  possible $Z$ decay channels. The dominant among these is of course the decay
into $q\bar q$ pairs (with a branching ratio of 
about $70\%$)\footnote{Here and in the
following, we imply a summation over all possible quark flavours
in the final states of all reactions we will consider.}, so that the most 
frequent HR signature  is 
four-jet final states. These are described by the
same diagrams as in Fig.~\ref{fig_Zgg}a, simply supplemented with 
an additional $Z\ar q\bar q$ decay current.
In the four-jet channel, now selected by applying the Durham algorithm
with cut-off $y_{\mathrm{cut}}=0.001$ to all {\sl four} partons
in the final state, one obtains the typical cross sections of 
Tab.~\ref{tab_Xsections}. (Note that we have also included the
$\gamma^*\ar q\bar q$ propagator in the $2\ar4$ HR matrix element.)
As already emphasised, the LEP2 four-jet rates are well below detection level. 
In contrast,
the NLC cross sections would still be detectable in the four-jet channel
after a few year of running at the level of 100 inverse femtobarns.

\begin{figure}[p]
\begin{center}
~\hskip-2.0cm\epsfig{file=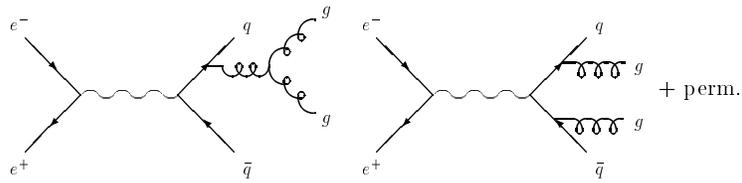,height=18cm}
\vskip-12.0cm
\caption{Representative Feynman diagrams contributing in
lowest order to \eeqqgg. 
An internal wavy line represents a $\gamma$ or a $Z$.
Permutations are not shown.}
\label{fig_tree}
\end{center}
\end{figure}

\begin{figure}[p]
\begin{center}
\vskip-5.0cm
~\epsfig{file=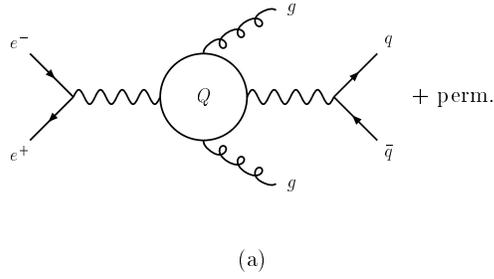,height=18cm}
\vskip-12.5cm
~\epsfig{file=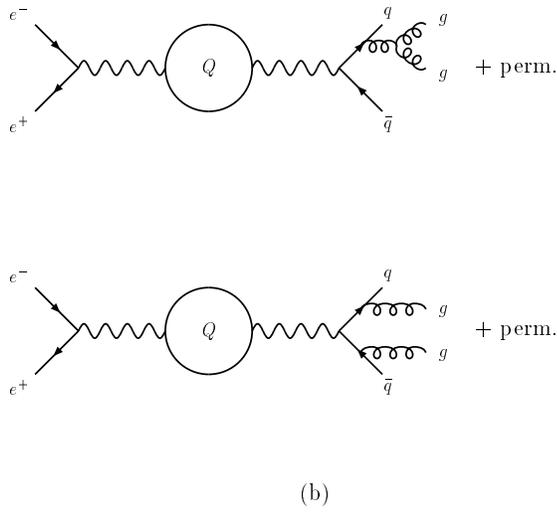,height=18cm}
\vskip-7.0cm
\caption{Representative Feynman diagrams contributing in
lowest order to the  process $e^+e^-\ar q\bar qgg$ via one quark loop
at order ${\cal O}(\alpha_{\mathrm{em}}^4\alpha_{\mathrm{s}}^2)$.
An internal wavy line represents a $\gamma$ or a $Z$.
Permutations are not shown.
Graphs (a) are the hadronic returns.
Graphs (b) are the $\gamma$/$Z$ self-energies.}
\label{fig_loop}
\end{center}
\end{figure}

However, a more interesting phenomenology could arise from the hadronic
returns in the four-jet channel  when they manifest themselves
in interference processes. In fact, the $q\bar q gg$ final state can also
be produced at tree-level by the ${\cal O}(\alpha_{\mathrm{em}}^2
\alpha_{\mathrm{s}}^2)$ diagrams of
Fig.~\ref{fig_tree}. These interfere with those in 
Fig.~\ref{fig_loop}a\footnote{To be more precise, only the Abelian
graphs do,  corresponding to the two gluons 
 produced in a colour singlet state.}.
This interference could allow for a more significant effect than the
square of the
HR amplitudes, possibly even  
at LEP2  energies, where the QCD process of Fig.~\ref{fig_tree}
is the dominant component of the total four-jet rate (even larger than
$e^+e^-\ar W^+W^-\ar q\bar q' Q\bar Q'$ at small values
of $y_{\mathrm{cut}}$). 

However, one should notice that the interference between the diagrams
in Fig.~\ref{fig_tree} and ~\ref{fig_loop}a is not the only
one occurring at this order of the couplings. 
In fact, one also has to consider the interference between the
tree-level graphs in Fig.~\ref{fig_tree} and  those in Fig.~\ref{fig_loop}b,
the latter being nothing more than the former supplemented with 
the $\gamma,Z$ self-energy at one-loop (and calculated assuming $\mu=\sqrt s$
for the renormalisation scale). 

The contributions of these two terms at LEP2 is presented in the upper
part of Tab.~\ref{tab_int},
alongside the leading term given by the square of the tree-level diagrams.
For reference, we have set the CM energy equal to 183~GeV. The
cross sections are calculated for two illustrative values of 
$y_{\mathrm{cut}}$ in the four-jet regime. From Tab.~\ref{tab_int} 
it can be seen
that the interference involving the HR amplitudes is positive
and very small, but still
 an order of magnitude larger than the square of the latter (see
Tab.~\ref{tab_Xsections}). In contrast, the interference involving the
self-energy diagrams is negative and sizeable, of the order of $2\%$
of the leading term. 
{}From these numbers, we conclude 
that  HR effects  are unobservable in practice
at LEP2 even in interference processes,. On the one hand,
their production rate is tiny {\sl per se}, and  on the other hand the 
other four-jet contributions 
(including the QCD \cite{noi} and EW \cite{Kleiss}
four-quark  $e^+e^-\ar q\bar q Q\bar Q$ components and their interference
\cite{pittau}) 
are much larger. In addition the situation is not improved by the 
next-to-leading order (NLO) QCD corrections to $e^+e^-\ar q\bar qgg$ and 
$e^+e^-\ar q\bar q Q\bar Q$ \cite{4jetnlo}, 
which give rise to a $K$-factor of order 1.5 \cite{a3} !  

\begin{figure}[p]
\begin{center}
~\epsfig{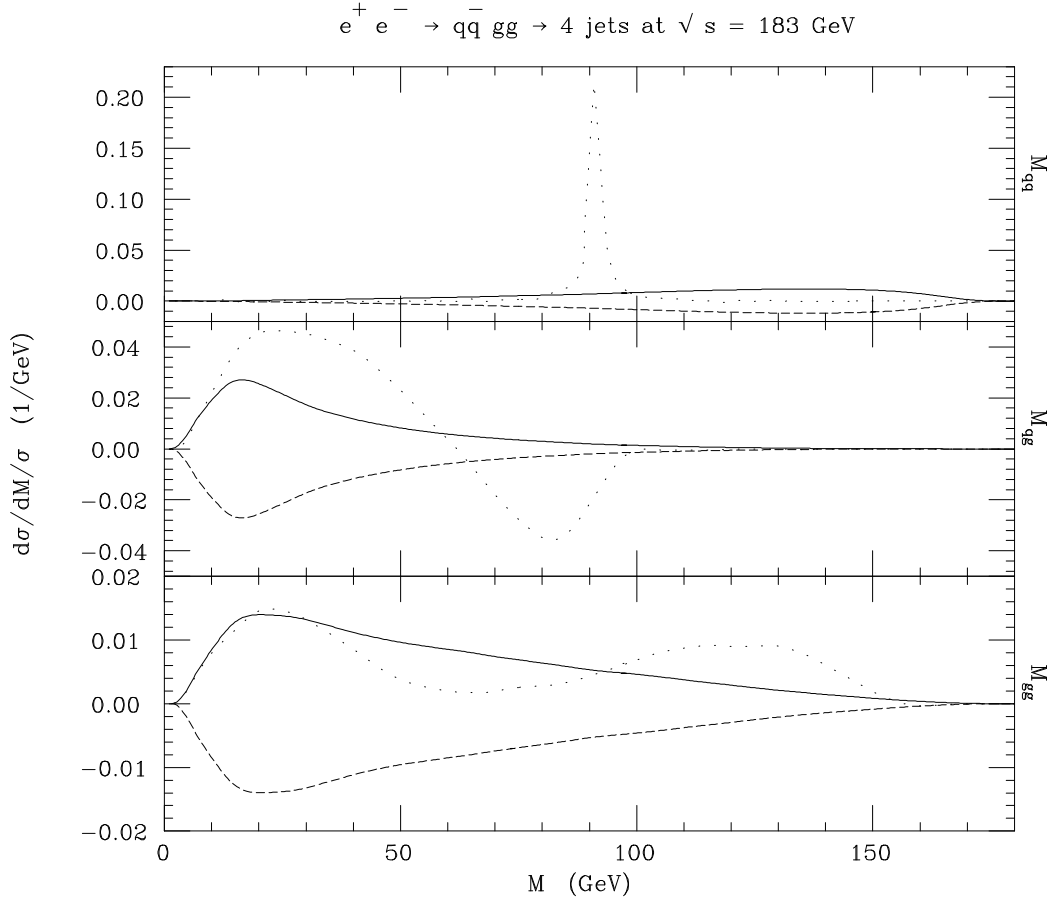}
\caption{Differential distributions in the invariant mass of the
$q\bar q$ (upper frame),
$qg$ (central frame) and
$gg$ (lower frame) pairs for  
the three sources of 
two-quark-two-gluon events defined  in the text:
$|M_{\mathrm{tree}}|^2$ (solid);
$2~{\mathrm{Real}}(M_{\mathrm{tree}} M_{\mathrm{self}}^*)$ (dashed);
$2~{\mathrm{Real}}(M_{\mathrm{tree}} M_{\mathrm{returns}}^*)$ (dotted). 
The CM energy is 183~GeV. The jet clustering algorithm 
used to separate four jets
is the Durham algorithm, with cut-off $y_{\mathrm{cut}}>0.001$. 
A summation over all possible
quark flavours in the final state
has been performed. 
Normalisation is to unity.}
\label{fig_m4j}
\end{center}
\end{figure}

Taking a more optimistic view, we can conclude that at LEP2 hadronic returns
do not constitute a serious background to the important physics 
measurements performed with the four-jet channel. For example,
the $W^+W^- \to\ $four jet process is used to measure $M_W$. 
Four-jet events can also be generated
by the elusive Higgs boson produced in association
with a $Z$, via $e^+e^-\ar ZH\ar q\bar q Q\bar Q$ (where the second
quark pair is dominantly $b\bar b$).  
Had the HR interference effects
been sizeable, it would have been a potentially serious  
background. In fact, by looking at Fig.~\ref{fig_m4j} one realises 
that not only the $q\bar q$ invariant mass induced by the diagrams
of Fig.~\ref{fig_loop}a resonates at $M_Z$\footnote{Further notice that
at $E_{\mathrm{cm}}=183$~GeV 
a detectable Higgs would have a mass $M_H$ approximately degenerate
with that of the $Z$ boson.}
but also the $qg$ (or, equivalently, $\bar qg$) mass has a (negative)
Jacobian peak around $M_W$ ! In contrast,  the $gg$ distribution
(lower plot)  would have been of little concern in this respect.
The behaviour of the other two four-jet mechanisms discussed is
driven by the infrared nature of the QCD interactions (small invariant masses
preferred) and by
the constraints enforced through the jet-finder, with 
the self-energy contributions being
naturally proportional to the tree-level ones.

Notice also that at an NLC with $E_{\mathrm{cm}}=350$~GeV
(see lower part of Tab.~\ref{tab_int}) hadronic returns do not 
enter the observable four-jet sample via interference effects. 
In fact, the corresponding rates are smaller in magnitude than those
produced by the HR amplitudes on their own (see Tab.~\ref{tab_Xsections}),
in contrast to the situation at LEP2. The sign is also reversed 
-- the $2~{\mathrm{Real}}(M_{\mathrm{tree}} M_{\mathrm{returns}}^*)$  
contributions are positive in the latter case, and negative in the former.

\begin{figure}[p]
\begin{center}
~\epsfig{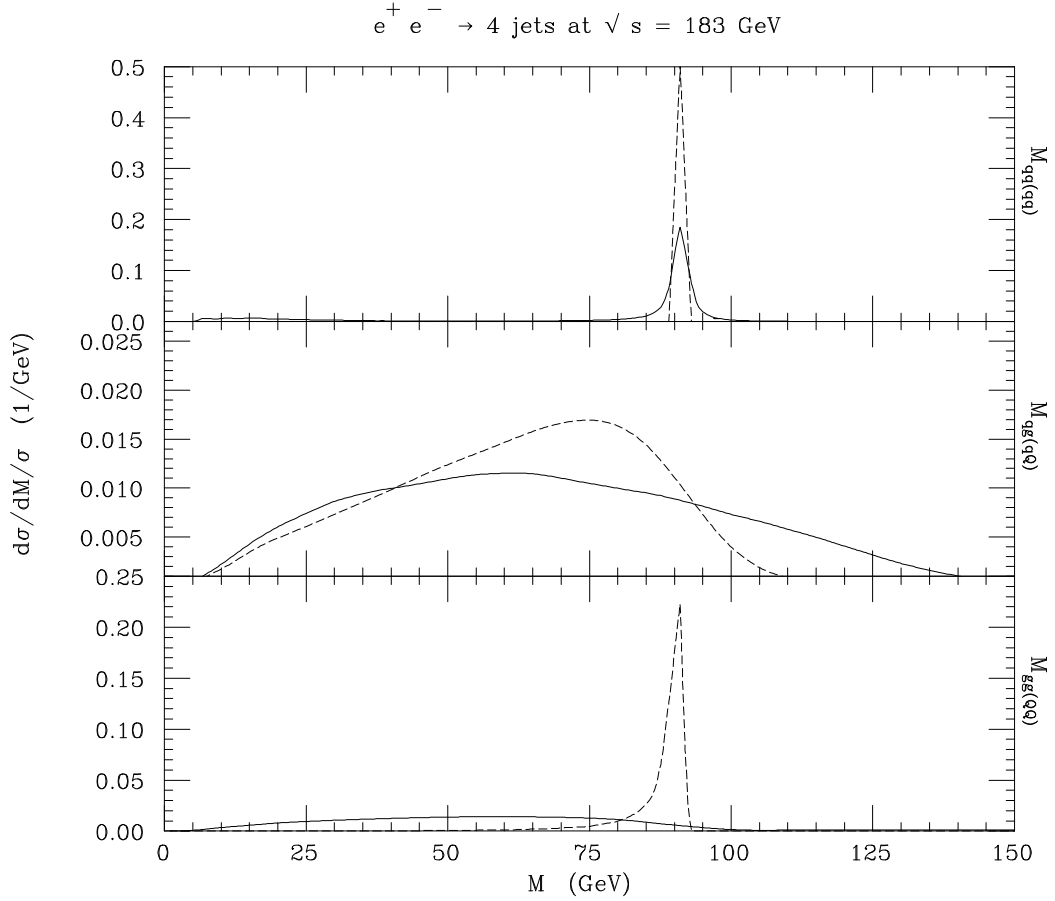}
\caption{Differential distributions in the invariant mass of the
$q\bar q(q\bar q)$ (upper frame),
$qg(q\bar Q)$ (central frame) and
$gg(Q\bar Q)$ (lower frame) pairs for  
the two processes: 
$e^+e^-\ar Z,\gamma^* gg \ar q\bar qgg$ (solid) and  
$e^+e^-\ar HZ \ar q\bar q Q\bar Q$ (dashed).
The CM energy is 183~GeV. The jet clustering algorithm 
used to separate four jets
is the Durham algorithm, with cut-off $y_{\mathrm{cut}}>0.001$. 
A summation over all possible
quark flavours in the final state
has been performed. 
Normalisation is to unity.}
\label{fig_square}
\end{center}
\end{figure}

Finally, for reference,
Fig.~\ref{fig_square} shows the same three mass distributions discussed
in the previous plots but now for the square of the
HR matrix element, i.e. $|M_{\mathrm{returns}}|^2$.  We again take
$E_{\mathrm{cm}}=183$~GeV and $y_{\mathrm{cut}}=0.001$ in the Durham algorithm.
For comparison, we also present in Fig.~\ref{fig_square} the invariant
mass of the $q\bar q$, $q\bar Q$ (or, equivalently, $\bar q Q$) and
$Q\bar Q$ pairs produced in the final state of the 
Higgs  process, when $M_H=M_Z$. 
Note the overlap of the degenerate Higgs and $Z$
boson mass peaks in the top frame, the latter also appearing 
in the bottom one (though cut off at the upper edge by phase space 
constraints on the $e^+e^-\ar ZH$ process) well above the spectrum 
generated by the squared diagrams of Fig.~\ref{fig_loop}a.
The two distributions in the central frame look rather similar.
For the $|M_{\mathrm{returns}}|^2$ contributions at
the NLC (which  we do not reproduce here), the $M_{gg}$ spectrum is basically
the same as that given in Fig.~\ref{fig_Zgg2}, the $M_{q\bar q}$ one
is again a Breit-Wigner centred around $M_Z$ (as in Fig.~\ref{fig_square}),
whereas the $M_{qg}$ one has a Jacobian shape extending  to values somewhat
higher than those in Fig.~\ref{fig_square}, having its maximum at about
100~GeV.

\section{Summary and conclusions}
\label{sec_conclusions}

We have studied in this paper the effects of what we
have called ``hadronic returns to the $Z$'' in high-energy $e^+e^-$
annihilation.
These correspond to diagrams in which a primary $\gamma,Z$ current
originates a quark loop from which two gluons are emitted in association
with a $Z$ boson, the former being energetic enough so that the
the latter is near its mass-shell. These are a novel source of 
two-lepton $+$ two-jet  and four-jet events which  could, {\sl a priori}, 
constitute  a problematic background for $W^+W^-$ and $ZH$ studies 
at LEP2 and NLC. Naive power counting arguments suggest 
that the rates could indeed be non-negligible. As there is no substitute
for a realistic analysis, we have performed a full matrix element 
calculation, which has enabled us to compare the HR cross sections with those 
of the standard processes. 
We first calculated the hadronic returns as a process on its own, by integrating
the amplitude squared of the relevant perturbative graphs, as well as 
the interference of  these with the tree-level
diagrams for quark-antiquark-two-gluon production, corresponding to the 
 case where the $Z$ boson decays into a quark pair.

In every channel studied, the effects of the HR contributions
 are completely negligible at LEP2.
In contrast, at an NLC with energy between 350 and 500~GeV, they can
be observed at a rate of a few per 100 inverse femtobarns for various
$Z$ decay signatures, whereas their interference with the tree-level
contribution to  the total four-jet rate is negligible also at the NLC.
Thus, the HR contributions are significant enough at such a
collider to merit attention when proceeding to experimental studies
of high luminosity data samples.

\section*{Acknowledgements}

SM would like to thank John Hill and the HEP Group at the Cavendish Laboratory
(Cambridge) for kindly allowing him to use their computers.
SM and DJM are grateful to the Centre for Particle Theory and Grey College
at Durham University for their kind hospitality. 
Discussions with Nigel Glover, Mike Seymour and 
Jay Watson are gratefully acknowledged.  
This work was supported in part by the EU IV Framework Programme
`Training and Mobility of Researchers', network `Quantum Chromodynamics
and the Deep Structure of Elementary Particles', 
contract FMRX-CT98-0194 (DG 12 - MIHT).

\vfill
\clearpage

\begin{table}[!t]
\begin{center}
\begin{tabular}{|c|c|c|c|c|c|}
\hline
\multicolumn{6}{|c|}
{\rule[0cm]{0cm}{0cm}
$\sigma(e^+e^-\ar \gamma^*/Z^* gg\ar q\bar q gg)$ (pb)} 
\\ \hline\hline
\multicolumn{6}{|c|}
{\rule[0cm]{0cm}{0cm}
$E_{\mathrm{cm}}$ (GeV)} 
\\ \hline\hline
$136$ & 
$161$ & 
$172$ & 
$183$ & 
$192$ & 
$205$ \\ \hline
$9.9\times10^{-7}$ & 
$1.3\times10^{-6}$ & 
$1.5\times10^{-6}$ & 
$1.8\times10^{-6}$ & 
$1.9\times10^{-6}$ & 
$2.1\times10^{-6}$ \\ \hline
\end{tabular}
\begin{tabular}{|c|c|}
\hline
$~~~~~~~~~~~~~~~~~~~~~350~~~~~~~~~~~~~~~~~~~~~$ & 
$~~~~~~~~~~~~~~~~~~~~~500~~~~~~~~~~~~~~~~~~~~~$ \\ \hline
$2.7\times10^{-6}$ & 
$2.1\times10^{-6}$ \\ \hline\hline
\multicolumn{2}{|c|}
{\rule[0cm]{0cm}{0cm}
Durham scheme \qquad \qquad \qquad \qquad $y_{\mathrm{cut}}>0.001$} 
\\ \hline
\end{tabular}
\vskip2.0cm
\caption{Cross sections for the hadronic returns
in the four-jet channel at eight different
energy points representative of the LEP2 (upper) and NLC (lower) colliders.
The jet clustering algorithm 
used to separate four jets
is the Durham algorithm, with cut-off parameter $y_{\mathrm{cut}}>0.001$. 
A summation over all possible
quark flavours in the final state
has been performed. The numerical errors do not
affect the significant digits shown.}
\label{tab_Xsections}
\end{center}
\end{table}

\vfill\clearpage

\begin{table}[!t]
\begin{center}
\begin{tabular}{|c||c|c|c|}
\hline
\multicolumn{4}{|c|}
{\rule[0cm]{0cm}{0cm}
$\sigma(e^+e^-\ar q\bar q gg)$ (pb)} 
\\ \hline\hline
\multicolumn{4}{|c|}
{\rule[0cm]{0cm}{0cm}
$E_{\mathrm{cm}}=183$ GeV} 
\\ \hline
$y_{\mathrm{cut}}$                                             &
$|M_{\mathrm{tree}}|^2$                                        &
$2~{\mathrm{Real}}(M_{\mathrm{tree}} M_{\mathrm{self}}^*)$     &
$2~{\mathrm{Real}}(M_{\mathrm{tree}} M_{\mathrm{returns}}^*)$  
\\ \hline
$0.001$     &
$+5.73$     & 
$-0.13$     & 
$+0.0000089$ 
\\
$0.010$     &
$+0.54$     & 
$-0.012$    & 
$+0.0000047$
\\ \hline\hline
\multicolumn{4}{|c|}
{\rule[0cm]{0cm}{0cm}
$E_{\mathrm{cm}}=350$ GeV} 
\\ \hline
$y_{\mathrm{cut}}$                                             &
$|M_{\mathrm{tree}}|^2$                                        &
$2~{\mathrm{Real}}(M_{\mathrm{tree}} M_{\mathrm{self}}^*)$     &
$2~{\mathrm{Real}}(M_{\mathrm{tree}} M_{\mathrm{returns}}^*)$  
\\ \hline
$0.001$     &
$1.09$     & 
$-0.020$     & 
$-0.00000019$ 
\\
$0.010$     &
$0.10$     & 
$-0.0019$    & 
$-0.00000084$
\\ \hline\hline
\multicolumn{4}{|c|}
{\rule[0cm]{0cm}{0cm}
Durham scheme} 
\\ \hline
\end{tabular}
\vskip2.0cm
\caption{Cross sections for the three sources of 
two-quark-two-gluon events defined  in the text:
$|M_{\mathrm{tree}}|^2$,
$2~{\mathrm{Real}}(M_{\mathrm{tree}} M_{\mathrm{self}}^*)$ and 
$2~{\mathrm{Real}}(M_{\mathrm{tree}} M_{\mathrm{returns}}^*)$, 
for two representative values of the cut-off $y_{\mathrm{cut}}$. 
The jet clustering algorithm 
used to separate the four jets
is the Durham algorithm. 
The CM energies are 183 and 350 GeV, as representative of LEP2 and NLC.
A summation over all possible
quark flavours in the final state
has been performed. The numerical errors do not
affect the significant digits shown.}
\label{tab_int}
\end{center}
\end{table}

\vfill\clearpage


\vfill\clearpage
 
\vfill\clearpage
 
\vfill\clearpage

\vfill\clearpage
 
\vfill\clearpage
 
\vfill\clearpage
 
\vfill\clearpage

\vfill\clearpage

\vfill\clearpage

\end{document}